\documentclass[prl,twocolumn,amsmath,amssymb,floatfix,showpacs,superscriptaddress]{revtex4}

\usepackage{graphicx}% Include figure files
\usepackage{dcolumn}% Align table columns on decimal point
\usepackage{times}
\usepackage{bm}% bold math

\begin{document}

\preprint{APS/123-QED}

\title{Anomalous interference in Aharonov-Bohm rings with two Majorana bound states}
 % Force line breaks with \\

 \author{Akiko Ueda}
 \email{akueda@hermes.esys.tsukuba.ac.jp}

 \affiliation{Division of Applied physics, Faculty of Pure and Applied Sciences, University of Tsukuba, Ibaraki, 305-8573, Japan}
 
\author{Takehito Yokoyama}
\affiliation{Department of Physics, Tokyo Institute of Technology, Tokyo 152-8551, Japan}

\date{\today}% It is always \today, today,
             %  but any date may be explicitly specified\UTF{00AB}%
\begin{abstract}
We investigate the conductance of an Aharonov-Bohm (AB) interferometer coupled to a quantum dot and two Majorana bound states on the edge of the topological superconductor with finite length. 
When the tunnel couplings between the Majorana bound states and the Aharonov-Bohm interferometer
are fixed to the specific phase, the differential conductance becomes zero 
irrespective of all the parameters as long as the hoppings to the two Majorana fermions on the opposite side are equal. 
The conductance at the zero bias voltage does not change with the magnetic flux penetrating the ring for all cases.
When the energy level of the quantum dot is equal to the energy of the Majorana bound states, the AB oscillation shows $\pi$ periodicity due to the particle-hole symmetry.   
The breaking of the time-reversal symmetry of the topological superconductor results in $2\pi$ periodicity of the AB oscillation for the specific phase
of the tunnel coupling while the time-reversal symmetry breaking leads to the mixing of the triplet and singlet states in the quantum dot in another specific phase. 
\end{abstract}

\maketitle

Exotic features of Majorana fermions \cite{Majorana} have been studied not only from the interest of the fundamental physics, 
but also from the application for quantum computing \cite{Wilczek,Beenakker,Alicea,Leijnse,Stanescu}.  
In recent years, Majorana fermions have been predicted in several setups with $s$-wave or $d$-wave superconductors \cite{Fu, Linder, Sau, Lutchyn}. It has been shown that Majorana fermions are manifested as a zero bias conductance peak in the normal metal/topological superconductor (TS) junctions \cite{Akhmerov,Tanaka, Law}. 
Recently, the zero bias conductance peak due to Majorana bound states (MBSs) has been observed in these systems \cite{Sasaki,Mourik,Deng,Das,Eduardo1}.
However, the signature is not enough to prove the existence of MBSs. 
The observed zero bias conductance peak is not quantized and it could occur even without MBSs in the presence of disorders \cite{Liu2,Pikulin,Bagrets,Kells,Eduardo2}.
Thus, alternative setups are necessary. 
Other setups are proposed to detect the phase information of transport electrons due to existence of the MBSs using 
interferometry of electron waves \cite{Liu, Benjamin}.

These studies mentioned above have examined the class D TSs, in which one Majorana fermion emerges at the edge.  
When the time-reversal symmetry is preserved, the TS is classified into the class BDI. Then, two Majorana fermions can exist on the edge of the TS which has the integer topological number of 2 \cite{Schnyder,Qi, He, Ii, Yamakage}. 
The BDI topological superconductors can be realized with superconductors coupled to the AIII topological insulators which show the quantum anomalous Hall effect\cite{Qi,He,Yamakage}.
Yamakage and Sato have reported that the zero bias conductance of the normal metal(N)-TS junction shows zero or $4e^2/h$ value depending on the phase of the tunnel coupling between the normal metal and the TS \cite{Yamakage}.
%when the couplings to the MBSs of the other side of the edge in TS are absent.

The Aharonov-Bohm (AB) ring with an embedded quantum dot attached to the normal lead has been studied in 2DEG in GaAs/AlGaAs heterostructures \cite{Kobayashi}. The high-order interference called the Fano resonance appears in the conductance as a function of the bias voltage when the high coherence is kept in the AB ring \cite{Fano}. The resonance peak shows an asymmetric structure with a peak and dip as a function of the bias voltage while AB oscillation shows $2\pi$ periodicity as a function of the magnetic flux penetrating the ring in such a system. 

In this paper, we consider an AB interferometer coupling to a quantum dot and two MBSs in the class BDI TS. 
%Here our interest is how the conductance is modified when one of the normal leads are replaced by the TS. 
We reveal that the zero bias conductance (which is always a peak or zero) is robust against the magnetic flux.
Also, the AB oscillation at the finite voltage shows $\pi$ periodicity under some conditions when the energy level of the quantum dot is equal to the energy of MBSs. Furthermore, when the phases of the tunnel couplings satisfy some conditions, the zero bias conductance vanishes, which is robust against any parameters except for the tunnel couplings to Majorana fermions at the opposite edge.  
We also calculate the anomalous Green function of the quantum dot to examine the proximity effect due to the MBSs which is reflected by the conductance. 

The schematic picture of the model is shown in Fig. 1. 
There are two paths between the leads N and TS, one path connects
the leads through a quantum dot, and the other path connects the leads directly by the reference arm.
The phase $\varphi$ represents the AB phase by the magnetic flux through the ring.
The chemical potential of the normal lead is  $\mu _{\rm N}=eV$. 
We adopt the effective Hamiltonian to describe the system. 
The Hamiltonian of the normal lead and the quantum dot are written as
\begin{align}
H_{N} = \sum_{k, \sigma} (\varepsilon_{k, \sigma} - \mu_{\rm N}) c^{\dagger}_{k \sigma}  c^{}_{k \sigma},
\end{align}
\begin{align}
H_{D} = \sum_{\sigma} \varepsilon_{0} c^{\dagger}_{0 \sigma} c^{}_{0 \sigma}.
\end{align}
Here, 
$c^{\dagger}_{k}$ and $c^{}_{k}$
denote the creation and annihilation operators of an electron of
momentum $k$ in the left lead. $\sigma$ denotes the spin. 
We assume that the energy spacing between the energy levels is larger than the energy broadening of one level in the dot. Thus, we consider a single energy level $\varepsilon_{0}$ in the dot
with creation and annihilation operators being $c_{0}^{\dagger}$ and $c^{}_{0}$, respectively.
In the TS with the topological number 2, there are two MBSs at each edge. 
The couplings between the Majorana zero energy bound states are  
\begin{align}
H_{M} &= \sum^{4}_{i<j} \bigl ( i\frac{E_{ij}}{2}\gamma_i \gamma_j - i \frac{E_{ji}}{2} \gamma_{j} \gamma_{i} \bigr), 
\end{align}
where $\gamma_{i}$ is the creation and annihilation operator for Majorana fermions labeled by  $i$ ($i=1,\cdots, 4$) and obeys $\gamma_{i}^2 = 1$.
The MBSs 1 (3) and 2 (4) are located in the same edge.  
When the time reversal symmetry is kept in the TS, the couplings between the same edge
($E_{12}$ and $E_{34}$) are prohibited \cite{He,Yamakage}. Moreover, only the couplings from one MBS to one of two MBSs at the other edge, $E_{13}$ and $E_{24}$, are permitted, while the couplings, $E_{14}$ and $E_{23}$ are prohibited.
%The topological superconductors which holds the relation above are classified into the BDI class topological superconductor \cite{He}. 
%When $E_{12}$ and $E_{34}$  are present, the degeneracy of the two MBSs is removed. 
Here, we consider the condition where the time-reversal symmetry is preserved ($E_{12}=E_{34}=0$) or not ($E_{12}$ or $E_{34}$ are not zero).
The tunnel couplings in the AB interferometer are expressed as
\begin{align}
H_{T}&=\sum_{k, \sigma} (t_{L} c^{\dagger}_{k \sigma} c^{}_{0 \sigma} + H. c.) \notag\\
& +\sum_{\sigma, i=1,2}( t^{\ast}_{R \sigma, i} c^{\dagger}_{0 \sigma}  - t_{R \sigma, i} c^{}_{0 \sigma}) \gamma_{i} \notag \\
&+ \sum_{\sigma, i=1,2} (W^{\ast}_{\sigma, i}e^{-i \varphi} c_{k \sigma}^{\dagger} - W_{\sigma, i}e^{ i \varphi}c^{}_{k \sigma}) \gamma_{i}. 
\end{align}
The tunnel couplings between electrons in the normal lead
and MBSs are spin-dependent even when the spin-orbit interaction is absent.

We calculate the current using the Keldysh Green function formalism \cite{Keldysh, Jauho},
 \begin{align}
 I_{\rm N} &= - e \langle \dot{N}_{\rm L, \sigma} \rangle
 \notag \\
 &= - ie\langle [H, N_{\rm L, }] \rangle
 \notag \\
 &=- \frac{e}{h} {\rm Re} \int d \omega \biggl [ t_{L} \sum_{k,\sigma} G^{<} _{ k\sigma, 0\sigma }(\omega) \notag \\
&+ \sum_{k, \sigma, i=1,2} W_{\sigma, i}e^{-i \varphi} G^{<}_{k\sigma, i}(\omega)  \biggr], 
\label{cur}
\end{align}
 where $N_{L} = \sum_{k, \sigma} c^{\dagger}_{k, \sigma} c_{k, \sigma}$.
Here,
$G^{<}_{k \sigma, 0 \sigma}(t,t^{\prime}) =i  \langle   c^{\dagger}_{0, \sigma }(t^{\prime}) c^{}_{k \sigma}(t)  \rangle$
and 
$G^{<}_{k \sigma, i}(t,t^{\prime}) = i  \langle   \gamma_{i}(t^{\prime}) c^{}_{k, \sigma}(t)  \rangle$.
We derive the Keldysh Green function $G^{<}_{k\sigma, 1} (\omega)$ and $G^{<}_{k\sigma, 0 \sigma}(\omega)$ using the equation of the motion formalism \cite{Jauho, Flensberg}. Note that the calculation is done without any approximation.

\begin{figure}
\includegraphics[width=8cm]{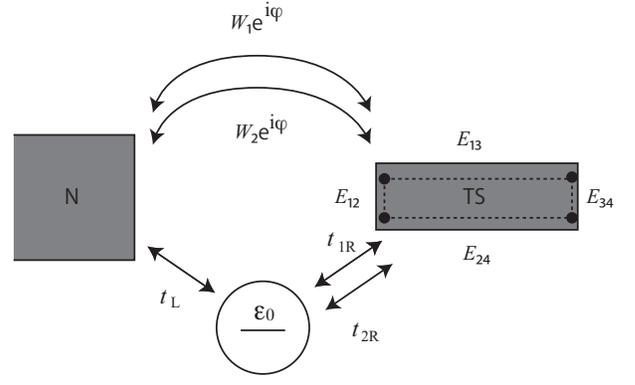}
\caption{Schematic setup of an Aharonov-Bohm interferometer with an embedded quantum dot coupling to TS.}
\label{f1}
\end{figure}

According to Ref. \onlinecite{Yamakage}, depending on the condition of the phases of the tunnel couplings, the system can be classified into the unitary and anti-unitary cases. The (anti-)unitary case is defined by $t_{\uparrow}=\eta t_{\downarrow}$ ($t_{\uparrow}=\eta t^*_{\downarrow}$ ) with $\eta=\pm 1$.
In the following, we consider the anti-unitary case and choose the tunnel coupling between the quantum dot and the MBSs as 
$t_{{\rm R} \uparrow, 1} = t e^{- i \pi/4}(-1)^{k}$,  $t_{{\rm R} \uparrow, 2} = t e^{ i \pi/4}(-1)^{l}$, 
$t_{{\rm R} \downarrow, 1} = t e^{i \pi/4}(-1)^{m}$, $t _{{\rm R} \downarrow, 2} = t  e^{- i \pi/4}(-1)^{n}$
where $k$, $l$, $m$ and $n$ denote integer numbers. The couplings between the normal lead and the MBSs are assumed to have the same relations, $W_{\uparrow, 1} = w e^{- i \pi/4}(-1)^{k}$,  $W_{\uparrow, 2} = w e^{ i \pi/4}(-1)^{l}$, 
$W_{ \downarrow, 1} = t e^{i \pi/4}(-1)^{m}$, $W_{ \downarrow, 2} = w e^{- i \pi/4}(-1)^{n}$. 
We define the energy broadening of the quantum dot due to the coupling to the normal lead as $\Gamma = \pi \nu t_{L}^2$ with the density of the states
of the normal lead $\nu$. 
Similarly, we define $\xi = \pi \nu w^2$. Below, we fix $\xi = t=\Gamma$.

We examine two cases, $(-1)^{n+m} = (-1)^{l + k}$ and  $(-1)^{n+m+1} = (-1)^{l + k}$.
When $(-1)^{n + m} = (-1)^{l + k}$ and the coupling to the MBSs  at the opposite edge is absent ($E_{13} = E_{24} = 0$), 
the maximum peak ($4e^2/h$) of the conductance appears at zero bias voltage when the energy level of a quantum dot is equal to the Majorana zero energy. 
In this paper, we call this phase of the tunnel coupling as the {\it constructive} phase.
When $(-1)^{n+m+1} = (-1)^{l + k}$, the conductance becomes always zero independent of the other parameters except for the tunnel couplings $E_{13}$ and $E_{24}$. Here we call this phase as the {\it destructive} phase.

\begin{figure}
\includegraphics[width=9cm]{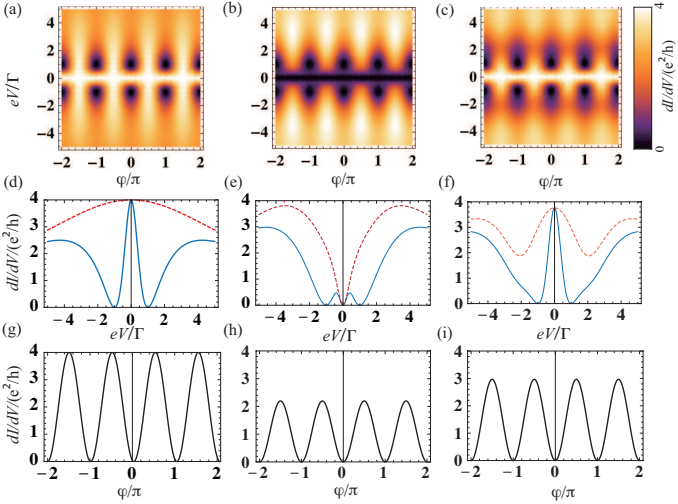}
\caption{(Color online) The differential conductance for the constructive phase [$(-1)^{l + k}= (-1)^{m + n}$]. We take $k= l= m= n=1$, $\xi =t=\Gamma$. (a) $dI/dV$ as a function of $\varphi$ and 
$eV$ for $E_{12} = E_{34} = E_{13}= E_{24} = 0$.
(b) $dI/dV$ as a function of $\varphi$ and 
$eV$ when $E_{12} = E_{34} = 0$ and $E_{13}/2= E_{24} = \Gamma$. (c) $dI/dV$ as a function of $\varphi$ and 
$eV$ when $ E_{12} = E_{34} = \Gamma$ and $E_{13}/2= E_{24} = \Gamma$. (d) $dI/dV$ as a function of $eV$  when $\varphi = 0$ (solid  line) and 
$\varphi = \pi/2$ (dotted line). The other parameters are the same as (a). (e) $dI/dV$ as a function of $eV$ when $\varphi = 0$ (solid  line) and 
$\varphi = \pi/2$ (dotted line). The other parameters are the same as (b).  (f) $dI/dV$ as a function of $eV$ when $\varphi = 0$ (solid  line) and 
$\varphi = \pi/2$ (dotted line). The other parameters are the same as (c).  (g) $dI/dV$ as a function of  $\varphi$ when $eV = \Gamma$.  The other parameters are the same as (a).  (h) $dI/dV$ as a function of $\varphi$ when $eV = \Gamma$. The other  parameters are the same as (b).  (i)  $dI/dV$ as a function of $\varphi$ when $eV = \Gamma$.  The other parameters are the same as (c). 
  }
\label{f2}
\end{figure}

\begin{figure}
\includegraphics[width=7cm]{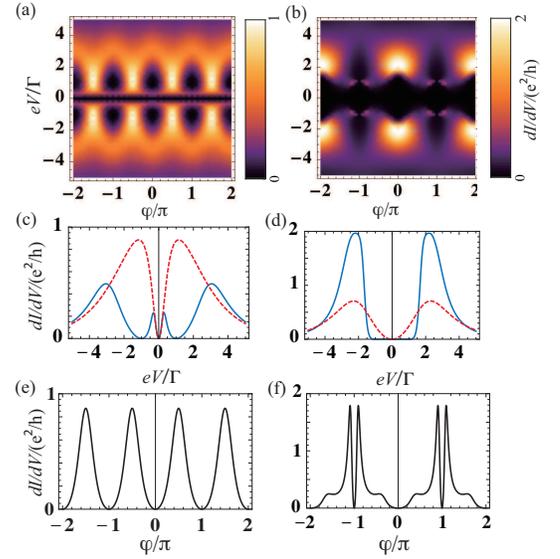}
\caption{(Color online) The differential conductance for the destructive phase [$(-1)^{l + k+1}= (-1)^{m + n}$].  We take $k=m=n = 1$ and $l=2$. $\xi =  t = \Gamma$.
(a) $dI/dV$ as a function of $\varphi$ and 
$eV$ for $E_{12} = E_{34}=0$, and $E_{13}/2= E_{24} = \Gamma$.
(b) $dI/dV$ as a function of $\varphi$ and 
$eV$ for $E_{12} = E_{34} = \Gamma$ and $E_{13}/2= E_{24} = \Gamma$.
(c) $dI/dV$ as a function of $eV$ for $\varphi = 0$ (solid  line) and $\varphi = \pi/2$ (dotted line). The other parameters are the same as (a).   
(d) $dI/dV$ as a function of $eV$ for $\varphi = 0$ (solid  line) and 
$\varphi = \pi/2$ (dotted line). The other parameters are the same as (b).  
(e) $dI/dV$ as a function of the magnetic phase $\varphi$ at $eV = \Gamma$. 
The other parameters are the same as (a). (f) $dI/dV$ as a function of $\varphi$ at $eV = \Gamma$. The other parameters are the same as (b). }
\label{f3}
\end{figure}

We start with the case of the constructive phase,  $(-1)^{l + k}= (-1)^{m + n}$. 
The energy level of the quantum dot is fixed to $\varepsilon_0 = 0$. 
The 2D plots in Fig. 2 indicate the $dI/dV$s as a function of the bias voltage $eV$ and the AB phase $\varphi$ 
for three cases, i.e. (1) all the couplings between MBSs are absent ($E_{12} = E_{34} = E_{13} = E_{24}=0$) [Fig. 2(a)], 
(2) the couplings between the MBSs on the opposite side are present  ($E_{12} = E_{34}=0$, $ E_{13}/2 = E_{24} = \Gamma$) [Fig. 2(b)], and
(3) all the couplings are finite ($E_{12} = E_{34}= E_{13}/2 = E_{24} = \Gamma$)[Fig. 2(c)].
%We can see $\pi$ periodicity for all the three cases. 
The zero bias conductance shows a peak for the case (1) and the case (3) but dip with zero at $eV=0$ for the case (2) with any $\varphi$.
To examine the result more precisely, the $dI/dV$ is plotted as a function of $eV$[Figs. 2(d), 2(e) and 2(f)].
For the case (1), the conductance reaches $4e^2/h$ at the zero bias voltage due to the resonant Andreev reflection through the MBSs, while
for the case (3), the zero bias conductance is less than $4e^2/h$. We have also calculated the conductance in the same setup with replacement of 
the TS by class D TS in which only one MBS locates at each edge (not shown in the figure). 
In this case, when the coupling to the MBS at the opposite side of the edge is absent, the zero bias peak is quantized as $2e^2/h$ (the maximum of the peak is half of that of the BDI TSs since the class D TSs have one MBS). However, when the coupling is present, the zero bias conductance shows a dip with zero value. The results in the case (1)[$E_{13} =E_{24}=0$] and (2)[$E_{13}/2=E_{24}\neq 0$] are similar  to the case of class D TS. 
However, the breaking of the time-reversal symmetry in the TS [case (3)]
suppresses the resonance (peak) and anti-resonance (dip).
%When all the hopping between Majorana bound states are absent, the peak and dip structure of the Fano resonance appears [Fig. 2 (d)].
The Fano resonance is symmetric due to the particle-hole symmetry for all the three cases. Note that Fano resonance becomes asymmetric when both leads are made of the normal metals in the present system. 
In Figs. 2(g), 2(h) and 2(i), AB oscillation exhibits the $\pi$ periodicity for all the three cases.
Note that the $\pi$ periodicity of the AB oscillation is also present within the superconducting gap when the topological superconducting lead is replaced by a topologically trivial superconducting one.

\begin{figure}
\includegraphics[width=8.5cm]{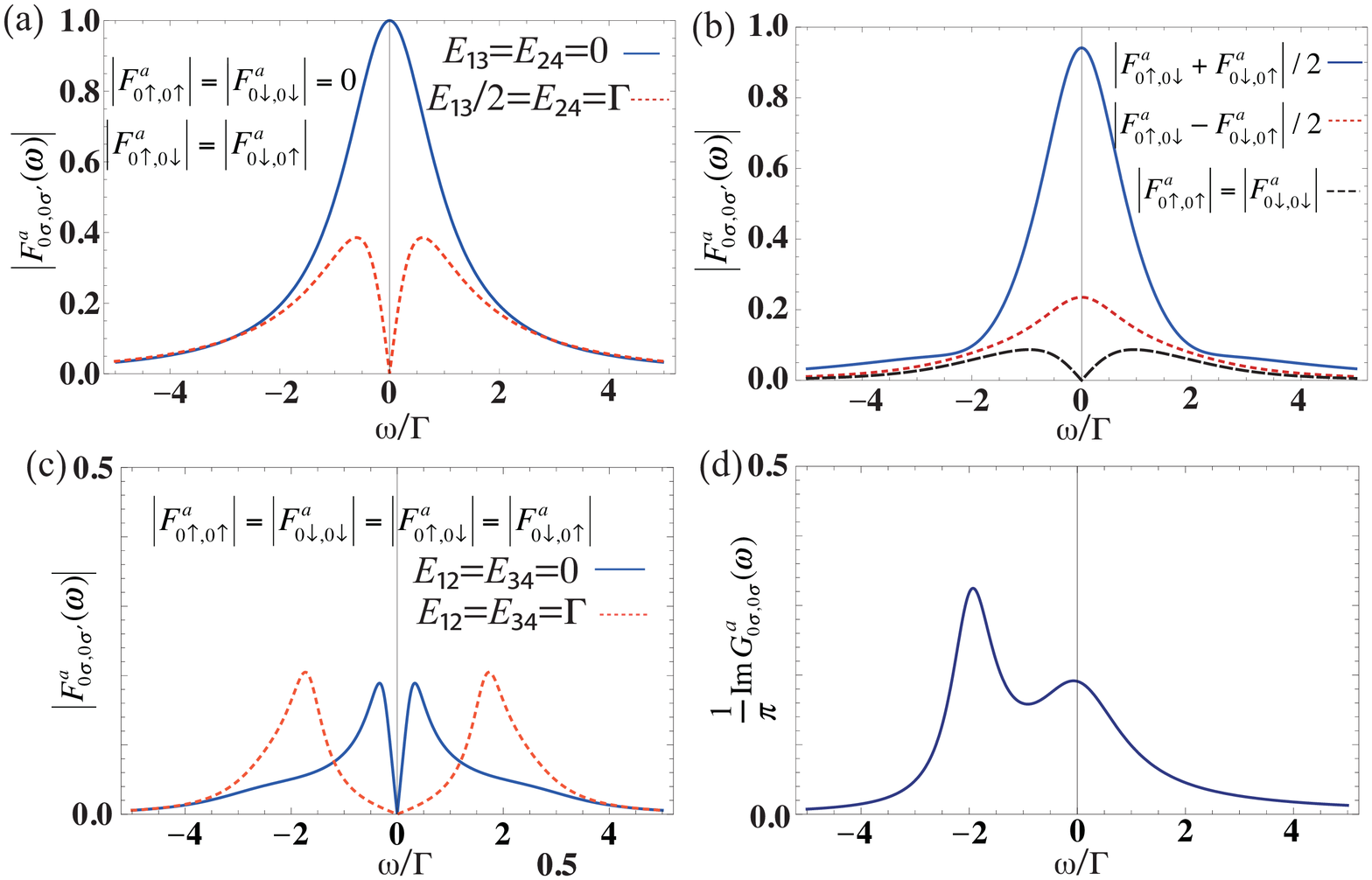}
\caption{(Color online) (a)-(c): The absolute value of the advanced anomalous Green function $|F^{a}_{0\sigma, 0\sigma^{\prime}}(\omega)|$ .  $\varphi =0$ for all cases. (a) $|F^{a}_{0 \uparrow, 0 \downarrow}|$ for the  constructive phase [$(-1)^{l + k+1}= (-1)^{m + n}$] when $E_{12} = E_{34} = E_{13}= E_{24} =0$ (solid line) 
and $E_{12} = E_{34}=0$, $E_{13}/2= E_{24} = \Gamma$ (dotted line). Here, $F^{a}_{0 \downarrow, 0 \uparrow}= F^{a}_{0 \uparrow, 0 \downarrow}$ and
$F^{a}_{0 \uparrow, 0 \uparrow} =F^{a}_{0 \downarrow, 0 \downarrow} = 0$. (b) $|F^{a}_{0 \uparrow, 0 \downarrow}(\omega)+F^{a}_{0 \downarrow, 0 \uparrow}(\omega)|/2$ (solid line), $|F^{a}_{0 \uparrow, 0 \downarrow}(\omega) - F^{a}_{0 \downarrow, 0 \uparrow}(\omega)|/2$ (dotted line), and 
$|F^{a}_{0 \uparrow, 0 \uparrow}(\omega)|$ (broken line) when $E_{12} = E_{34}=E_{13}/2= E_{24} = \Gamma$ for the constructive phase.  
$F^{a}_{0 \uparrow, 0 \uparrow} =- F^{a}_{0 \downarrow, 0 \downarrow}$.
(c)  $|F^{a}_{0 \uparrow, 0 \downarrow}(\omega)|$ for $E_{13}/2= E_{24} = \Gamma$ and  $E_{12}= E_{34} = 0$ (solid line) or $E_{12}= E_{34} = \Gamma$
(dotted line) for the destructive phase. 
$|F^{a}_{0 \uparrow, 0 \downarrow}| = |F^{a}_{0 \downarrow, 0 \uparrow}| = |F^{a}_{0 \uparrow, 0 \uparrow}|=|F^{a}_{0 \downarrow, 0 \downarrow}|$. 
(d) The energy spectrum in the quantum dot $\frac{1}{\pi}{\rm Im} G^{a}_{0\sigma, 0\sigma} (\omega)$ for the destructive phase with $E_{13}= E_{24} = E_{12}= E_{34} = 0$ when $\varphi =0$.}
\label{f4}
\end{figure}

Next, we consider the case of destructive phase in Fig. 3.
%The couplings between the MBSs are fixed to the same parameters with the three cases in Fig. 2.
Interestingly, we find $dI/dV = 0$ independent of $eV$ and $\varphi$
when $E_{13} = E_{24}$ (not shown in a figure since the conductance is always zero as a function of $eV$ and $\varphi$).
Note that the vanishing of the conductance is robust against the energy level of the quantum dot and the other parameters except for $E_{13}$ and $E_{24}$.
When $E_{13} \neq E_{24}$,  the $dI/dV$ recovers nonzero value [Fig. 3(a) and 3(b)]. However,
the conductance is always zero at $eV = 0$ [Fig. 3(c) and 3(d)] while, in the constructive phase, the zero bias conductance ranges from zero to $4e^2/h$ depending on the couplings between MBSs. 
Also, the conductance as a function of $eV$ shows a gap in the vicinity of zero bias voltage.
When the time-reversal symmetry is broken ($E_{12} \neq 0$ or $E_{34} \neq 0$), the period of the AB oscillation changes to 
$2 \pi$ from $\pi$ [Fig. 3(e) and 3(f)].
This result is in stark contrast to that in the constructive phase since the breaking of the time-reversal symmetry does not affect the $\pi$ periodicity in the constructive phase. 
To investigate the origin of the $2\pi$ periodicity in the destructive phase with the broken time reversal symmetry, 
we calculate the lesser Green's function of Majorana fermions $G^{<}_{i,i}(t, t^{\prime})=i \langle \gamma_{i}(t^{\prime}) \gamma_{i}(t) \rangle$ ($i=1,2$).
The number of MBSs is calculated as $n_{i}=- i/\pi \int d \omega G^{<}_{i,i}(\omega)$.  
We find that $n_{i}(eV) = n_{i}(-eV)$ ($i = 1$ and 2)  for the constructive phase while $n_{i}(eV) \neq n_{i}(-eV)$ for at least one of the MBSs ($i = 1$ or 2) for the destructive phase. Thus, the results show that the particle-hole symmetry of the MBSs is broken only for the destructive phase when the time-reversal symmetry 
is absent. Therefore, the breaking of the particle-hole symmetry is the origin of the $2 \pi$ periodicity.
Let us point out that the $\pi$ periodicity is realized only when $\varepsilon_{0} = 0$.  Note that when the energy level of the quantum dot is not equal to zero, we see $2 \pi $ periodicity for both constructive and destructive phases even when the time-reversal symmetry is preserved since the particle-hole symmetry is broken.

To examine these results more precisely, we calculate the anomalous Green function in the dot 
$F^{a}_{0\sigma, 0\sigma^{\prime}}(t, t^{\prime}) = i \theta(t^{\prime} - t) \langle \{ c^{\dagger}_{0 \sigma}(t),  c^{\dagger}_{0 \sigma^{\prime}} (t^{\prime}) \}\rangle$. 
In Fig. 4, we plot the absolute value of $F^{a}_{0\sigma, 0\sigma^{\prime}}(\omega)$. In the constructive phase [Fig. 4 (a)], when the time reversal symmetry is preserved, the absolute value of $F^{a}_{0\uparrow, 0\downarrow}(\omega) [= F^{a}_{0\downarrow, 0\uparrow}(\omega)]$ has a maximum at $\omega = 0$ for $E_{13}=E_{24}=0$ (solid line) and becomes zero  at $\omega = 0$ for $E_{13}/2=E_{24} = \Gamma$ (dotted line) which correspond to the values of conductance at $eV = 0$ for both cases. Note that the conductance has the additional dip and peak structures due to the Fano resonance. When the time reversal symmetry is absent [Fig. 4 (b)], the singlet state characterized by $[F^{a}_{0\downarrow, 0\uparrow}(\omega) - F^{a}_{0\downarrow, 0\uparrow}(\omega)]/2$ (dotted line) appears in addition to the triplet state $[F^{a}_{0\downarrow, 0\uparrow}(\omega) + F^{a}_{0\downarrow, 0\uparrow}(\omega)]/2$ (solid line). 
For the destructive phase [Fig. 4(c)], the value of the conductance at $eV = 0$ also corresponds to the absolute value of $F^{a}_{0\sigma, 0\sigma^{\prime}}(\omega)$
at $\omega=0$ which is zero for the both cases, $E_{12} = E_{34} = 0$ (solid line) and $E_{12}=E_{24} =\Gamma$ (dotted line). 
The relation between the couplings of the MBSs and the anomalous Green function is summarized in Table I.

In the constructive phase,  when time reversal symmetry is kept, the Cooper pairs induced by the proximity effect inside the AB interferometer are purely triplet. 
Note that the orbital symmetry of the induced pairing is isotropic $s$-wave in the dot. Thus, the induced pairing is of odd-frequency triplet symmetry \cite{Berezinskii,Tanaka2}. By analytical calculations, we obtain the anomalous Green's functions in the dot as $F^{a}_{0\uparrow, 0\uparrow}(\omega)  \propto (E_{13}^2 - E_{24}^2) (t^2+ \Gamma \xi e^{2i \varphi})eV$.
%When $E_{12}$ or $E_{34}$ is present where the time-reversal symmetry is not preserved, the Cooper pair shows mixed state of triplet and singlet symmetries. 

In the destructive phase, we find $F^{a}_{\sigma, \sigma^{\prime}}=0$ when $E_{13} = E_{24}$. Therefore, the conductance becomes zero independent of $eV$ and $\varphi$.
Note that the vanishing of the conductance is robust against the shift of the energy level of the quantum dot $\varepsilon_{0}$. The density of states in the quantum dot $1/\pi {\rm Im} G^{a}_{0\sigma,0\sigma}(\omega)$ is shown in Fig 4 (d). It is found that even though the density of states in the quantum dot is finite, the conductance is zero in the destructive phase.
When $E_{13}\neq E_{24}$, the absolute values of the anomalous Green function of all the spins are the same, and also the relations $F^{a}_{0\uparrow, 0\downarrow} = F^{a}_{0 \downarrow, 0\uparrow}$ and $F^{a}_{0\uparrow, 0\uparrow} = - F^{a}_{0\downarrow, 0\downarrow} $ are satisfied. The Cooper pairs owing to the MBSs are always purely triplet even when the time reversal symmetry is broken inside the TS ($E_{12} \neq 0$ or $E_{34} \neq 0$), which is different from the case of the constructive phase. 

In this paper, we have considered the anti-unitary case.
We have also investigated the unitary case: $t_{R\uparrow, i}= \eta t_{R \downarrow, i}$ and $W_{\uparrow, i}= \eta W_{\downarrow, i}$. In the unitary case, the couplings for $(-1)^{n + m} = (-1)^{l + k}$ are classified to the destructive phase while those for $(-1)^{n + m} = (-1)^{l + k +1}$ are classified to the constructive phase.\cite{Yamakage}
We have found that the behaviors of the conductance in each phase are similar to those in the anti-unitary case.

\begin{widetext}
\begin{table}[h]
\centering
\begin{tabular}{|c|c|c|c|}
\hline
& $E_{13}=E_{24}$, $E_{12} = E_{34} = 0$ & $E_{13}\neq E_{24}$, $E_{12} = E_{34} = 0 $ & $E_{13}\neq E_{24}$, $E_{12} \neq 0$ or  $E_{34} \neq 0 $ \\
\hline
{\it constructive phase} & $ F^{a}_{0\uparrow, 0\downarrow} = F^{a}_{0\downarrow, 0\uparrow}$, $ F^{a}_{0\uparrow, 0\uparrow} = F^{a}_{0\downarrow, 0\downarrow} =0 $ & 
$ F^{a}_{0\uparrow, 0\downarrow} = F^{a}_{0\downarrow, 0\uparrow}$, $F^{a}_{0\uparrow, 0\uparrow} = - F^{a}_{0\downarrow, 0\downarrow}$
  &$ |F^{a}_{0\uparrow, 0\downarrow}| \neq| F^{a}_{0\downarrow, 0\uparrow}|$, $F^{a}_{0\uparrow, 0\uparrow} = - F^{a}_{0\downarrow, 0\downarrow}$\\
\hline
{\it distructive phase} &  $ F^{a}_{0\uparrow, 0\downarrow} = F^{a}_{0\downarrow, 0\uparrow}= F^{a}_{0\uparrow, 0\uparrow} = F^{a}_{0\downarrow, 0\downarrow} =0 $ & $F^{a}_{0\uparrow, 0\downarrow} =F^{a}_{0 \downarrow, 0\uparrow}$, $F^{a}_{0\uparrow, 0\uparrow} = - F^{a}_{0\downarrow, 0\downarrow} $ &
$F^{a}_{0\uparrow, 0\downarrow} = F^{a}_{0 \downarrow, 0\uparrow}$, $F^{a}_{0\uparrow, 0\uparrow} = - F^{a}_{0\downarrow, 0\downarrow} $\\
\hline
\end{tabular}
\caption{The relation between the anomalous Green functions in the quantum dot $F^{a}_{0 \sigma, 0\sigma^{\prime}}(\omega)$ and various coupling parameters between the MBSs.} 
\end{table}
\end{widetext}

In conclusion, we have investigated the conductance of an AB interferometer coupled to a quantum dot and two MBSs at each edge of the TS. Fano resonance is symmetric as a function of the bias voltage when the energy level of the quantum dot is equal to the energy level of the MBSs. 
The zero bias conductance is not affected by the magnetic flux penetrating the ring.
The AB oscillation at the finite bias voltage holds $\pi$ periodicity when the time-reversal symmetry is preserved.
When the tunnel couplings between the interferometer and MBSs are fixed to the specific phase and the hoppings to the two Majorana fermions at the opposite side are equal, the differential conductance become zero independent of all parameters. 
When the time reversal symmetry is kept, the Cooper pair induced by the proximity effect is purely triplet both 
for the constructive and the destructive phases. When the coupling between the two MBSs in the same edge is allowed, 
the Cooper pair in the dot becomes a mixture of triplet and singlet states for the constructive phase (but the AB oscillation still shows $\pi$ periodicity) while the destructive phase holds the purely triplet state, but the period of AB oscillation changes to $2 \pi$. 

This work was supported by Grant-in-Aid for Young Scientists (B) (No. 23740236, 24710111) and the ``Topological Quantum Phenomena" (No. 25103709) Grant-in Aid for Scientific Research on Innovative Areas from the Ministry of Education, Culture, Sports, Science and Technology (MEXT) of Japan.

\appendix


\begin{thebibliography}{9}
%\bibitem{golub} 
%A. Golub, I. Kuzmenko, and Y. Avishai, Phys. Rev. Lett. {\bf 107}, 176802 (2011).
\bibitem{Majorana} E. Majorana: Nuovo Cimento \textbf{14} 171 (1937).

\bibitem{Wilczek} F. Wilczek: Nat. Phys. \textbf{5} (2009).

\bibitem{Alicea} J. Alicea: Rep. Prog. Phys. \textbf{75} 076501 (2012).

\bibitem{Beenakker} C. W. J. Beenakker:  Annu. Rev. Con. Mat. Phys. \textbf{4} 113 (2013).

\bibitem{Leijnse} M. Leijnse and K. Flensberg, Semicond. Sci. Technol. \textbf{27}, 124003 (2012).

\bibitem{Stanescu} T. D. Stanescu and S. Tewari, J. Phys. Condens. Matter \textbf{25}, 233201 (2013).


%\bibitem{Sato} M. Sato: Phys. Lett. B \textbf{575}126 (2003).

\bibitem{Fu} L. Fu and C. L. Kane: Phys. Rev. Lett. \textbf{100} 096407 (2008).
%\bibitem{Fu2} L. Fu and C. L. Kane, Phys. Rev. Lett. \textbf{102}, 216403 (2009).

\bibitem{Sau} J. D. Sau, R. M. Lutchyn, S. Tewari, and S. Das Sarma: Phys. Rev. Lett. \textbf{104} 040502 (2010).

\bibitem{Linder} J. Linder, Y. Tanaka, T. Yokoyama, A. Sudb{\o}, and N. Nagaosa: Phys. Rev. Lett. \textbf{104} 067001 (2010); Phys. Rev. B \textbf{81} 184525 (2010). 

\bibitem{Lutchyn} R. M. Lutchyn, J. D. Sau, and S. Das Sarma: Phys. Rev. Lett. \textbf{105} 077001 (2010).

\bibitem{Akhmerov} A. R. Akhmerov, J. Nilsson, and C. W. J. Beenakker: Phys. Rev. Lett. \textbf{102} 216404 (2009). 

\bibitem{Tanaka} Y. Tanaka, T. Yokoyama, and N. Nagaosa: Phys. Rev. Lett.
\textbf{103} 107002 (2009).

\bibitem{Law} K. T. Law, P. A. Lee, and T. K. Ng: Phys. Rev. Lett. \textbf{103} 237001 (2009).

\bibitem{Sasaki} S. Sasaki, M. Kriener, K. Segawa, K. Yada, Y. Tanaka, M. Sato, and Y. Ando: Phys. Rev. Lett. \textbf{107} 217001 (2011).

\bibitem{Mourik} V. Mourik, K. Zuo, S. M. Frolov, S. R. Plissard, E. P. A. M. Bakkers, and L. P. Kouwenhoven: Science \textbf{336} 1003 (2012).

\bibitem{Deng} M. T. Deng, C. L. Yu, G. Y. Huang, M. Larsson, P. Caroff, and H. Q. Xu, Nano Lett. \textbf{12}, 6414 (2012).

%\bibitem{Rokhinson} L. P. Rokhinson, X. Liu, and J. K. Furdyna, Nat. Phys. \textbf{8}, 795 (2012).

\bibitem{Das} A. Das, Y. Ronen, Y. Most, Y. Oreg, M. Heiblum, and H. Shtrikman, Nat. Phys. \textbf{8}, 887 (2012). 

\bibitem{Eduardo1} Eduardo J. H. Lee, Xiaocheng Jiang, Manuel Houzet, Ram\"{o}n Aguado, Charles M. Lieber, and Silvano De Franceschi, Nat. Nanotech.  {\textbf 9}, 79 (2014).

\bibitem{Liu2} J. Liu, A. C. Potter, K. T. Law, and P. A. Lee, Phys. Rev. Lett. \textbf{109}, 267002 (2012).

\bibitem{Pikulin} D. I. Pikulin, J. P. Dahlhaus, M. Wimmer, H. Schomerus and C. W. J. Beenakker, New J. Phys. \textbf{14}, 125011 (2012).

\bibitem{Bagrets} D. Bagrets and A. Altland, Phys. Rev. Lett. \textbf{109}, 227005 (2012).

\bibitem{Kells} G. Kells, D. Meidan, and P. W. Brouwer, Phys. Rev. B \textbf{86}, 100503 (2012).

\bibitem{Eduardo2} Eduardo J. H. Lee, Xiaocheng Jiang, Ram\"{o}n Aguado, Georgios Katsaros, Charles M. Lieber, and Silvano De Franceschi, Phys. Rev. Lett.  {\textbf 109}, 186802 (2012).

\bibitem{Liu} D. E. Liu and H. U. Baranger: Phys. Rev. B \textbf{84} 201308(R) (2011).

\bibitem{Benjamin} C. Benjamin and J. K. Pachos, Phys. Rev. B {\textbf 81}, 085101 (2010).

%\bibitem{Hutzen} 
%R. Hutzen, A. Zazunov, B. Braunecker, A. Levy Yeyati, and R, Egger, 
%Phys. Rev. Lett. {\bf 109}, 166403 (2012). 

\bibitem{Schnyder}
A. P. Schnyder, S. Ryu, A. Furusaki, and A. W. W. Ludwig, Phys. Rev. B {\bf 78}, 195125 (2008). 

\bibitem{Qi}
X.-L. Qi, T. L. Hughes, and S.-C. Zhang, Phys. Rev. B {\bf 82}, 184516 (2010). 

%X.-L. Qi, T. L. Hughes, S. Raghu, and S.-C. Zhang, Phys. Rev. Lett. {\bf102}, 187001 (2009).

\bibitem{Ii}
A. Ii, A. Yamakage, K. Yada, M. Sato and Y. Tanaka, 
Phys. Rev. B {\bf 86}, 174512 (2012).

\bibitem{He} J. J. He, J. Wu. T-P Choy, X-J Liu, Y. Tanaka and K. T. Law,  Nat. Commun. {\bf 5}, 3232 (2014).

\bibitem{Yamakage} A. Yamakage and M. Sato, Physica E 55, 13 (2014)

\bibitem{Kobayashi}
K.\ Kobayashi, H.\ Aikawa, S.\ Katsumoto, and Y.\ Iye,
Phys.\ Rev.\ Lett.\ {\bf 88}, 256806 (2002);
Phys.\ Rev.\ B {\bf 68}, 235304 (2003).

\bibitem{Fano}
U.\ Fano, Phys.\ Rev.\ {\bf 124}, 1866 (1961).

\bibitem{Keldysh}
L.\ V.\ Keldysh, Zh.\ Ekps.\ Teor.\ Fiz.\ {\bf 47}, 1515 (1965).

\bibitem{Jauho}
A.\-P Jauho, N.\ S.\ Wingreen, and Y.\ Meir, Phys.\ Rev.\ B {\bf 50}, 5528 (1994).

\bibitem{Flensberg} 
K. Flensberg, Phys. Rev. B {\bf 82}, 180516(R) (2010).


\bibitem{Berezinskii} V. L. Berezinskii, JETP Lett. {\bf 20}, 287 (1974). 

\bibitem{Tanaka2} Y. Tanaka, M. Sato, and N. Nagaosa, J. Phys. Soc. Jpn. {\bf 81}, 011013 (2012). 

\end{thebibliography}
\end{document}